\documentclass[twocolumn, 10pt, prb, aps, showpacs, reprint, amsmath, amssymb, superscriptaddress]{revtex4-1}

\usepackage[pdftex]{graphicx}

\newcommand{\sinc}{\mathrm{sinc}}
\newcommand{\nSi}{n_\mathrm{Si}}
\newcommand{\nSiO}{n_\mathrm{SiO_2}}

\begin{document}

\title{Three-dimensional emission from organic Fabry-Perot microlasers}
\author{Cl\'ement Lafargue}
\email{clement.lafargue@ens-cachan.fr}
\author{Stefan Bittner}
\author{Sergii Lozenko}
\author{Joseph Lautru}
\author{Joseph Zyss}
\affiliation{Laboratoire de Photonique Quantique et Mol\'eculaire, CNRS UMR 8537, Institut d'Alembert FR 3242, Ecole Normale Sup\'erieure de Cachan, 61 Avenue du Pr\'esident Wilson, F-94235 Cachan, France}
\author{Christian Ulysse}
\affiliation{Laboratoire de Photonique et Nanostructures, CNRS UPR20, Route de Nozay, F-91460 Marcoussis, France}
\author{Christophe Cluzel}
\affiliation{Laboratoire de m\'ecanique et technologie Cachan, ENS-Cachan / CNRS UMR 8535 / Paris VI University / UniverSud Paris PRES, 61 avenue du Pr\'esident Wilson, F-94235 Cachan, France}
\author{M\'elanie Lebental}
\affiliation{Laboratoire de Photonique Quantique et Mol\'eculaire, CNRS UMR 8537, Institut d'Alembert FR 3242, Ecole Normale Sup\'erieure de Cachan, 61 Avenue du Pr\'esident Wilson, F-94235 Cachan, France}

\date{\today}

\begin{abstract}
We measured the far-field emission patterns in three dimensions of flat organic dye microlasers using a solid angle scanner. Polymer-based microcavities of ribbon shape (i.e., Fabry-Perot type) were investigated. Out of plane emission from the cavities was observed, with significant differences for the two cases of resonators either fully supported by the substrate or sustained by a pedestal. In both cases, the emission diagrams are accounted for by a model combining diffraction at the cavity edges and reflections from the substrate.
\end{abstract}

\maketitle

\noindent The integration of optical technologies and microelectronics is one of the key areas in current photonics research \cite{Matsko2009}, in particular regarding micro-sources. 
An important aspect for applications is the directionality of the emission from microlasers (see for instance Ref.\cite{yu2012beam}).
Many experiments with and simulations of flat microresonators with a thickness much smaller than their horizontal extension investigate only the emission in the plane of the cavity \cite{Gmachl1998,Harayama2003,Chern2003, Kurdoglyan2004, ben2005unidirectional, Wiersig2006a, Lebental2007a, Song2011a}.
Such microresonators can be easily fabricated on chip using lithography techniques \cite{Lebental2006}.
Since, however, the thickness of the resonators is of the order of the wavelength, diffraction occurs at the edges of the vertical side walls, possibly leading to emission out of the plane of the cavities.
This has seldom been investigated experimentally \cite{Peter2007, Kim2007, Lee1998, renner2006whispering} or theoretically \cite{ikegami1972reflectivity, vukovic1999facet}, however.
In order to measure the emission in all three dimensions, we have constructed a solid angle scanner \cite{Kim2007}, a setup allowing for the variation of both the azimuthal and polar angle of observation. 
We used it to investigate organic solid-state microlasers based on dye-doped polymers which can be fabricated easily down to nanoscale resolution \cite{lebental2009organic}.
Their spectral and emission properties have been intensively investigated\cite{bogomolny2011trace,Schwefel2004,Gozhyk2012} and depend sensitively on the cavity shape.
Here we used ribbon-shaped Fabry-Perot (FP) cavities that are one of the simplest types of microlasers. 
Their 3D far-field emission was measured  and interpreted by an analytical model\cite{Peter2007,Lee1998} that is simpler than previously proposed ones\cite{ikegami1972reflectivity}.
Reflections at the substrate add complexity to the emission patterns\cite{Harayama2012,Peter2007}. 
Therefore experiments were carried out with cavities either supported by a pedestal \cite{Lozenko2012} or fully in contact with the substrate.
In both cases, a good agreement with the model is obtained, which can be further extended to other resonator shapes.

\begin{figure}[tb]
\begin{center}
\includegraphics[width = 6.5 cm]{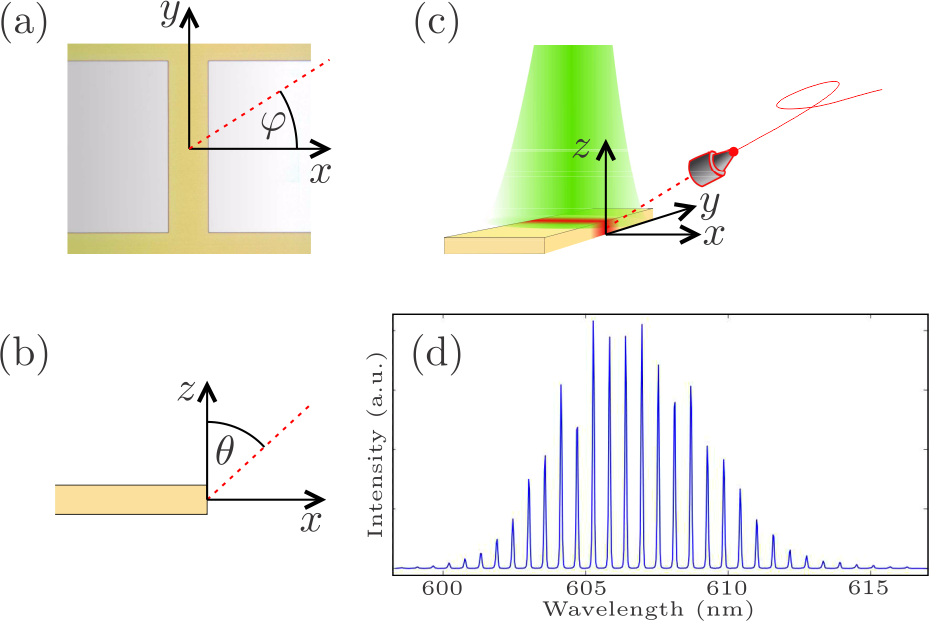}
\end{center}
\caption{(color online) (a) Top view of a ribbon microlaser (photograph from an optical microscope in real colors). The dye-doped polymer layer appears yellow. (b) Schematic side view of the microlaser. (c) Oblique sketch (not to scale) of the microlaser including the pump beam (green), the laser emission (red) and the detection lens. (d) Typical lasing spectrum emitted by a ribbon microlaser.}
\label{fig:cavGeom}
\end{figure}

The microlasers were fabricated as follows. A thin layer ($\sim 700$ nm) of poly(methylmethacrylate) (PMMA) do\-ped with 5\% wt.\ of the laser dye DCM\footnote{4-(Dicyanomethylene)-2-methyl-6-(4-di\-methyl\-amino\-styryl)-4\textit{H}-pyran} is spin-coated on a silicon wafer with a $2~\mu$m thick layer of silica. The cavity structures are then created using electron beam lithography, which ensures vertical sidewalls and sharp edges with sub-wavelength accuracy. A photo of a ribbon-shaped cavity with a width of $200~\mu$m is shown in Fig.~\ref{fig:cavGeom}(a). In order to produce cavities supported by a pedestal, the samples were dipped in hydrofluoric acid so as to partly under-etch the silica layer beneath the polymer cavities without damaging the organic part \cite{Lozenko2012}.

\begin{figure}[tb]
\begin{center}
\includegraphics[width = 6.0 cm]{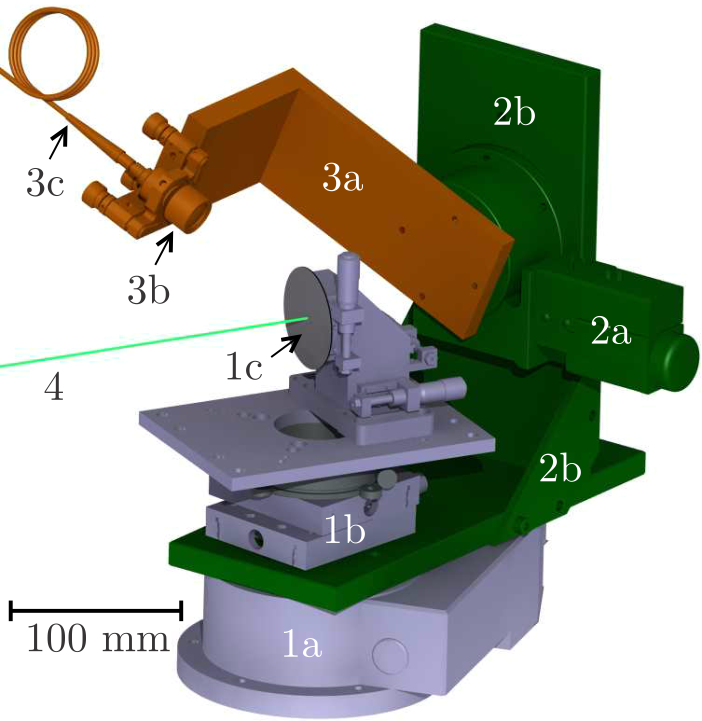}
\end{center}
\caption{(color online) Drawing of the detection setup. Parts 1 (grey) are fixed. Parts 2 (green) are in rotation around a vertical axis. Parts 3 (orange) are in rotation with respect to parts 2 around a horizontal axis. 1a, 2a --  motorized rotation stages. 1b -- translational stages to position the sample in the center of rotation. 1c -- wafer with microlasers. 2b, 3a -- rotating arms. 3b -- collection lens. 3c -- optical fiber. 4 -- pump beam.}
\label{fig:sasDesign}
\end{figure}

The microlasers were pumped just above their lasing threshold by a frequency-doubled Nd:YAG laser ($532$~nm, $500$~ps, $10$~Hz) with linear polarization parallel to the y axis [see Fig.~\ref{fig:cavGeom} (c) for notations and Ref.\cite{Gozhyk2012}]. The emission from the cavity into arbitrary directions was collected with a lensed fiber and transferred to a spectrometer. A typical measured spectrum is shown in Fig.~\ref{fig:cavGeom}(d). The sample holder itself does not move with respect to the table, whereas the collection lens is positioned by two rotating arms as shown in Fig.~\ref{fig:sasDesign}. To ensure that the cavity stays aligned with the lens, the two motors rotating the arms must have a very small eccentricity ($\pm 2~\mu$m for Newport RV120PP rotation stages), and their axes must cross each other with a mismatch tolerance of the order of $10~\mu$m. The latter was achieved via precise manufacturing of the arms and assembling them with centering pins. The wafer with the microlasers was attached vertically to the setup so that the horizontal pump beam is perpendicular to the wafer.

\begin{figure}[tb]
\begin{center}
\includegraphics[width = 7 cm]{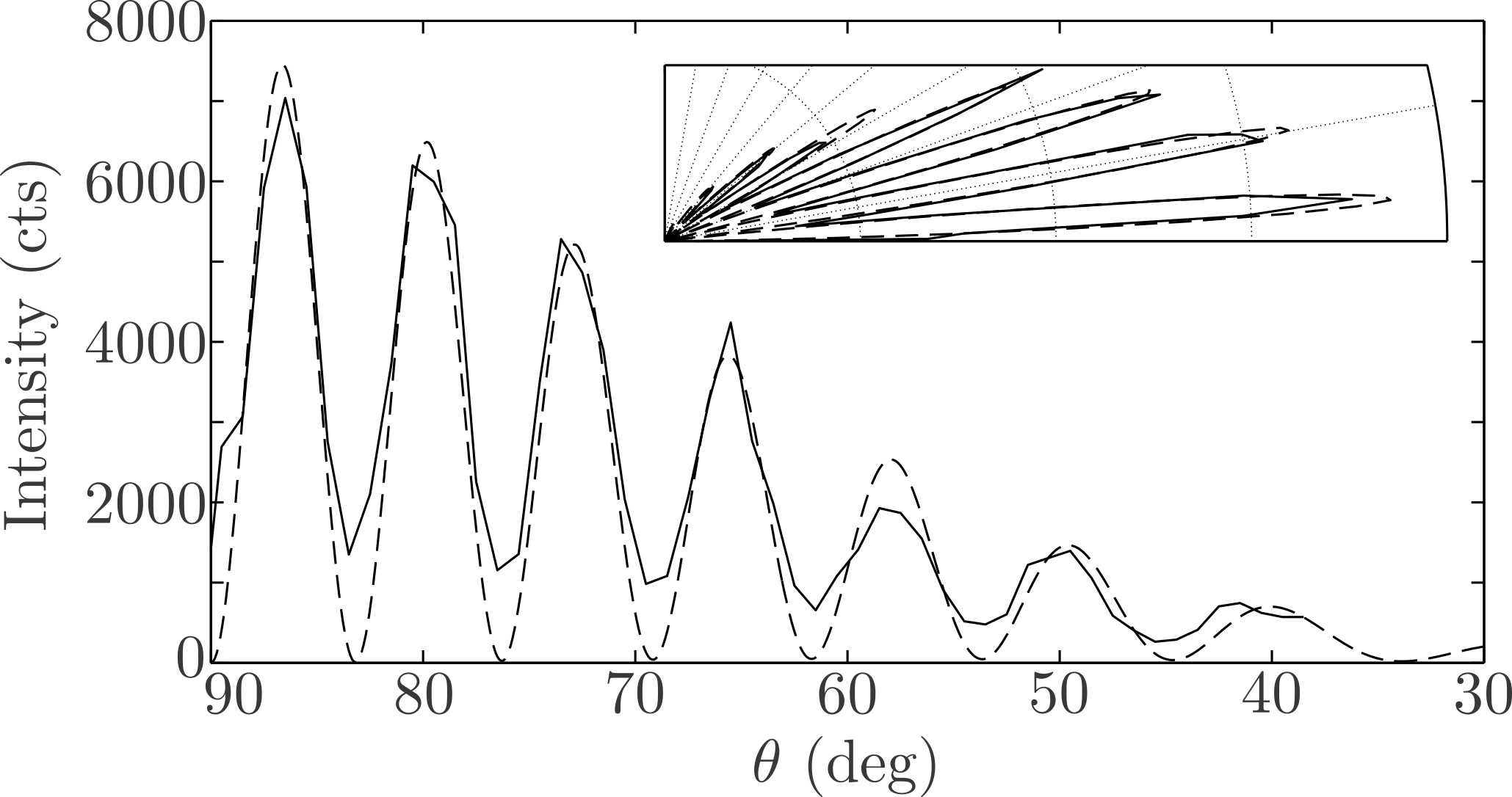}
\end{center}
\caption{Latitude emission diagram for a resonance at $605$~nm of a FP microlaser on a pedestal. Note the inverted $\theta$ axis. The solid line is the measured data, and the dashed line a fit of the model [Eqs.~(\ref{eq:factorization})--(\ref{eq:diffPattern})]. The inset shows the emission diagram in polar coordinates.}
\label{fig:farfieldPed}
\end{figure}

Previous studies demonstrated that many properties of flat microlasers like their spectra can be explained in the framework of a 2D effective refractive index approximation \cite{Lebental2007}. Just above threshold, ribbon (FP) cavities exhibit only one kind of lasing modes that is based on the bouncing-ball orbit along the $x$ direction. Feedback along the $y$ direction is suppressed because photons travelling in that direction are lost in the polymer layer [cf.~Fig.~\ref{fig:cavGeom}(a)]. Consequently, FP cavities emit only in the $\varphi = 0^\circ$ and $\varphi = 180^\circ$ directions \cite{Lebental2007}. Therefore, we measured the latitude emission diagram, that is the far-field lasing intensity with respect to the polar angle $\theta$ at fixed $\varphi = 0^\circ$.
The latitude emission diagram for a cavity on a pedestal is shown in Fig.~\ref{fig:farfieldPed}.
There is no significant dependence on the angle $\theta$ of the spectral and polarization features of the emission, unlike the cases reported in Ref.\cite{Kim2007}.
The major part of the lasing emission is not mainly in the cavity plane ($\theta = 90^\circ$) as could have been expected, but mostly out of the plane of the cavity. Moreover, the emission diagram exhibits an oscillating pattern that does not depend on the FP width.

\begin{figure}[tb]
\begin{center}
\includegraphics[width = 5 cm]{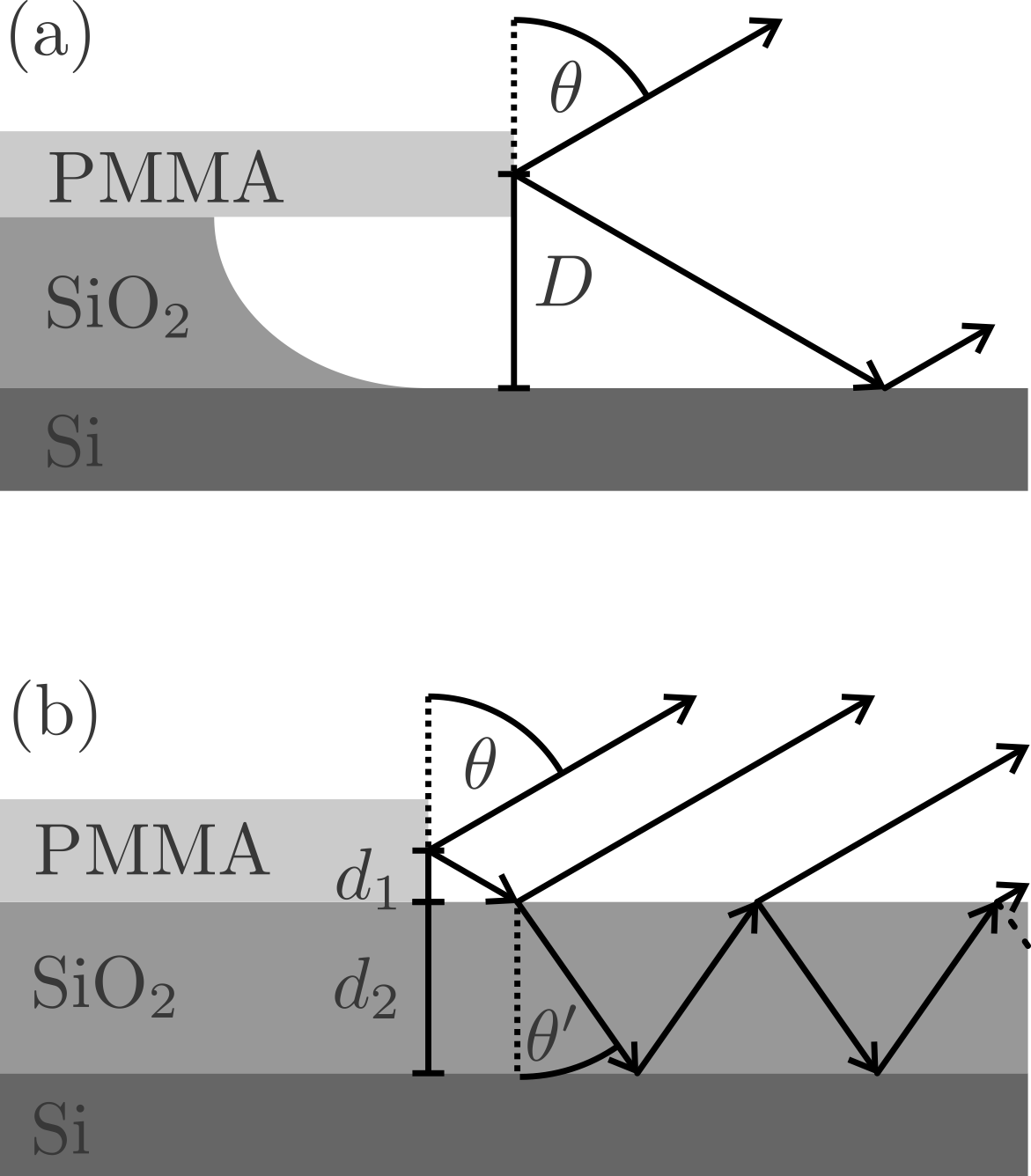}
\end{center}
\caption{Side views of the cavity geometry (not to scale) and interfering ray trajectories. (a) Cavity on a pedestal. (b) Cavity fully sustained by the silica substrate.}
\label{fig:rayGeom}
\end{figure}

In fact, the observed emission pattern $I(\theta)$ originates from two different effects, namely reflections at the substrate and diffraction at the cavity edges. It can therefore be expressed as a product of two terms \cite{Peter2007},
\begin{equation} \label{eq:factorization} I(\theta) = R(\theta) \cdot F(\theta) \, . \end{equation}
The factor $R(\theta)$ results from the interference between the direct ray trajectory from the sidewall to the lens and the one reflected at the silicion wafer as shown in Fig.~\ref{fig:rayGeom}(a).
It has the form \cite{Peter2007}
\begin{equation} \label{eq:reflFacPedestal} R(\theta) = \left| 1 + r(\theta) e^{-2 i k D \cos{\theta}} \right|^2 \, , \end{equation}
where $D$ is the distance between the silicon surface and the center of the polymer layer [see Fig.~\ref{fig:rayGeom}(a)], $k$ is the wave number, and $r(\theta)$ is the Fresnel reflection coefficient at the air/silicon interface for s polarization since the FP microlaser emits light polarized in the cavity plane \cite{Gozhyk2012}. This term is responsible for the oscillatory behavior that dominates the emission pattern. The second factor, $F(\theta)$, is related to the diffraction at the edges and determines the envelope of the emission pattern. It is the Fourier transform (FT) of the electric near-field distribution $f(z)$ at the cavity side wall \cite{Lee1998}, which is, however, not known \textit{a priori}. For the sake of simplicity it is assumed that the near field $f(z)$ is homogeneous over a certain length $l$ which is of the order of the cavity thickness. The resulting diffraction pattern is therefore that of a slit, namely, 
\begin{equation} \label{eq:diffPattern} F(\theta) \propto \sinc^2 \left( \frac{k l \cos{\theta}}{2} \right) \, . \end{equation}
This factor provides the envelope of the latitude diagram whose behavior is dominated by the oscillatory factor $R(\theta)$. Therefore, the exact shape of $F(\theta)$ cannot be determined here. 

A fit of this model is plotted as dashed line in Fig.~\ref{fig:farfieldPed} and shows very good agreement. The fitted parameters are $D_\mathrm{fit} = 2.55~\mu$m and $l_\mathrm{fit} = 556$~nm, where $\nSi = 3.94$ was used for the refractive index of silicon. The thicknesses of the polymer and silica layers were measured with a profilometer to be $700$~nm and $2.1~\mu$m, respectively. As expected, the value of $l_\mathrm{fit}$ is of the order of the thickness of the polymer layer. The measured thicknesses correspond to a value of $D_\mathrm{exp} = 2.45~\mu$m, from which the fitted value deviates by less than $4\%$.

\begin{figure}[tb]
\begin{center}
\includegraphics[width=7cm]{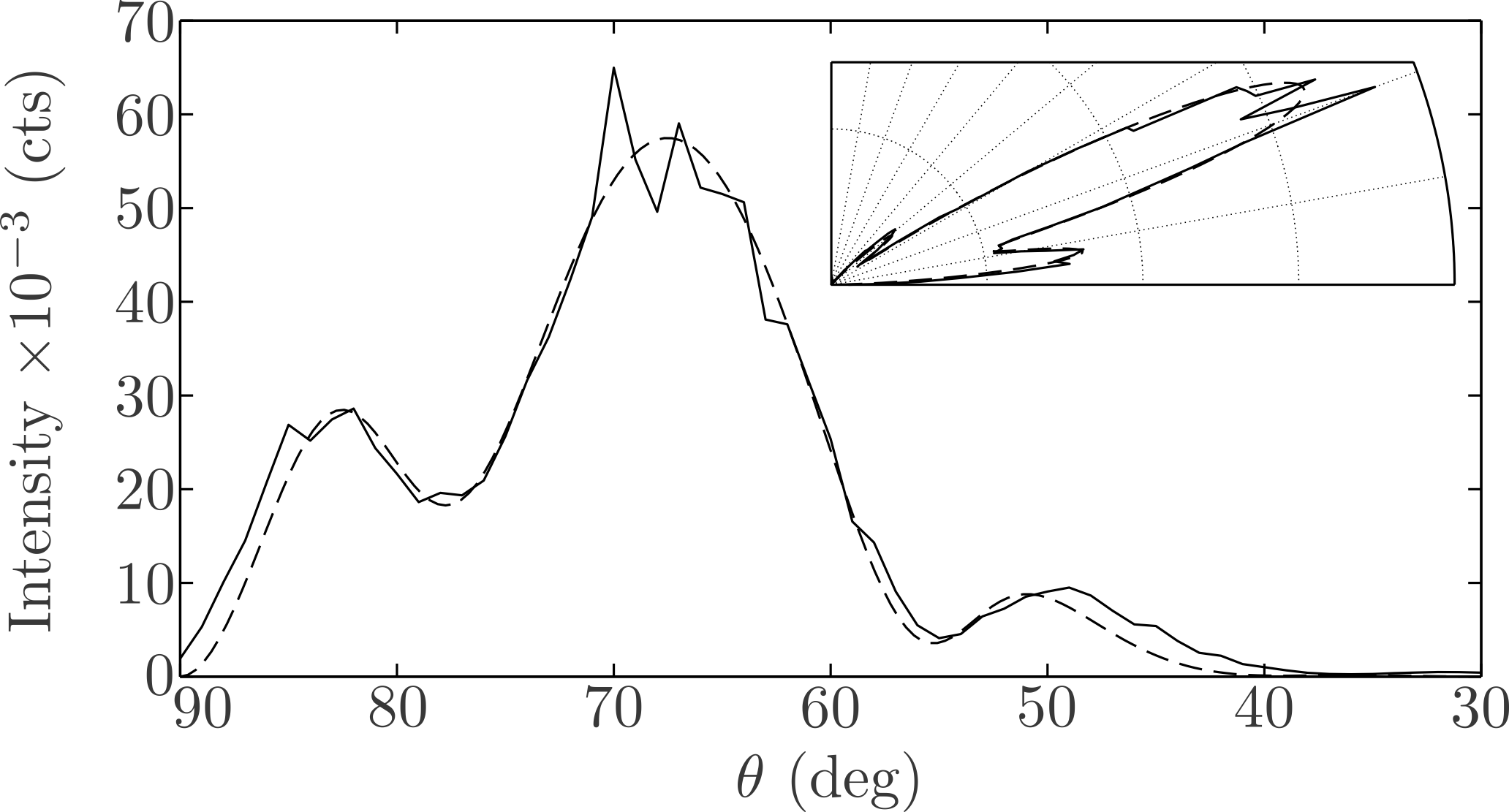}
\end{center}
\caption{Latitude emission diagram for a FP microlaser fully sustained by the substrate. The solid line is the measured data, and the dashed line a fit of the model [Eqs.~(\ref{eq:factorization}), (\ref{eq:diffPattern}) and (\ref{eq:reflFacSubstrate})]. The inset shows the emission diagram in polar coordinates.}
\label{fig:farfieldSub}
\end{figure}

The latitude emission diagram for a FP cavity fully supported by the silica coated substrate is shown in Fig.~\ref{fig:farfieldSub}. 
The main emission direction is obviously out of the plane and concentrated around $\theta = 65^\circ$ with two smaller lateral lobes around $80^\circ$ and $50^\circ$. The interference pattern $R(\theta)$ is more complicated here since the direct trajectory interferes with one reflected at the air-silica interface and an infinite number of reflections at the silica-silicon interface as depicted in Fig.~\ref{fig:rayGeom}(b). Adding up all trajectories results in 
\begin{equation}  \label{eq:reflFacSubstrate} 
\begin{array}{rl}
R(\theta) = & \Big| 1 + r_\mathrm{I}(\theta) e^{-i k \delta_1}  \\
 & + t_\mathrm{I}(\theta) t_\mathrm{II}(\theta') \frac{e^{-i k (\delta_1 + \delta_2)} r_\mathrm{III}(\theta')}{1 - e^{-ik \delta_2} r_\mathrm{III}(\theta') r_\mathrm{II}(\theta')} \Big|^2 \, , 
\end{array} 
\end{equation}
with $\delta_1 = 2 d_1 \cos{\theta}$, $\delta_2 = 2 d_2 [\nSiO - \sin{\theta} \sin{\theta'}] / \cos{\theta'}$, and $\sin{\theta'} = \sin{\theta} / \nSiO$ with $\nSiO = 1.46$. 
The Fresnel reflection and transmission coefficients for angle of incidence $\theta$ and s polarization are denoted by $r_\mathrm{I, II,III }(\theta)$ and $t_\mathrm{I, II,III }(\theta)$, respectively, where I stands for the air/silica, II for the silica/air and III for the silica/silicon interface.
The fit parameters are $d_1$, which is half the thickness of the cavity, $d_2$, which is the thickness of the silica layer [see Fig.~\ref{fig:rayGeom}(b)], and $l$. The fit (dashed line in Fig.~\ref{fig:farfieldSub}) shows very good agreement with the measurement for parameters $d_{1, \mathrm{fit}} = 347$~nm, $d_{2, \mathrm{fit}} = 2.118~\mu$m, and $l_\mathrm{fit} = 697$~nm. It should be noted that the shape of the latitude emission diagram depends sensitively on the parameters $d_1$ and $d_2$. The measured thickness of the PMMA layer is $2 d_{1, \mathrm{exp}} = 680$~nm, which agrees well with the fitted values for $d_1$ and $l$. The thickness of the silica layer, $d_{2, \mathrm{exp}} \approx 2.1~\mu$m like for the pedestal sample, also agrees well with $d_{2, \mathrm{fit}}$. 

In summary, we have investigated flat organic microlasers and observed emission out of the cavity plane with a solid angle scanner. For ribbon-shaped cavities, the measured latitude emission diagrams were compared to a simple analytical model taking into account diffraction at the cavity edges and reflections from the substrate, showing excellent agreement. 
The methodology applied here can be generalized to other, more complicated cavity shapes, and allow an insight into the mode structure inside the resonator, since the diffraction patterns $F(\theta)$ are related to the electric field distributions at the vertical cavity side walls. These cannot be predicted analytically but contain important information on the cavity modes. 
They are important to understand the reflection of waves at such interfaces with finite height \cite{ikegami1972reflectivity, vukovic1999facet}. For instance, the reflection coefficients for finite interfaces are of great relevance for the determination of lasing thresholds and to improve on the effective refractive index approximation \cite{Bittner2009}. In addition, the diffraction pattern is directly connected to the diffraction at dielectric corners and edges, which remains an open problem \cite{Gennarelli2011}.

\vspace{2 mm}
S.~B. gratefully acknowledges funding from the European Union Seventh Framework Programme (FP7/2007-2013) under grant agreement n$^\circ$ 246.556. The authors thank S.~Colin for fruitful discussions regarding the design of the solid angle scanner, and M.~Boudreau and G.~Bader for ellipsometric measurements of refractive indices.

\end{document}